\documentclass[showpacs,aps,prc,twocolumn,nofootinbib]{revtex4-2}
\usepackage{graphicx}
\usepackage{xcolor}
\usepackage{amsmath}
\usepackage{comment}
\usepackage{hyperref}
\usepackage{lineno}
\usepackage{scrextend}
\usepackage[export]{adjustbox}

\newcommand{\epem}{e$^+$e$^-$}
\newcommand{\epemmath}{e^+e^-}
\newcommand{\mee}{$M_{\rm{e}^+\rm{e}^-}$}
\newcommand{\meemath}{M_{\rm{e}^+\rm{e}^-}}
\newcommand{\mllmath}{M_{\ell^+\ell^-}}
\newcommand{\mrho}{$M_{\rho}$}
\newcommand{\mrhomath}{M_{\rho}}

\newcommand{\NN}{$N+N$}
\newcommand{\mmiss}{$M_{\rm{miss}}$}
\newcommand{\pt}{$p_{\rm{T}}$}
\newcommand{\qdeux}{$q^2$}
\newcommand{\piz}{$\pi^0$}

\newcommand{\pim}{$\pi^-$}
\newcommand{\pimp}{\pim\ + p}
\newcommand{\pippim}{$\pi^+\pi^-$}
\newcommand{\pippimmath}{\pi^+\pi^-}
\newcommand{\piN}{$\pi$+N}
\newcommand{\pimC}{\pim\ + C}
\newcommand{\pNb}{p + Nb}
\newcommand{\pp}{p + p}
\newcommand{\pimCHdeux}{\pim\ + CH$_2$}
\newcommand{\chdeux}{CH$_2$}
\newcommand{\pimptonee}{\pim p $\to$ n\epem }
\newcommand{\pimptonpippim}{\pim p $\to$ n\pippim }
\newcommand{\pimpnrho}{\pim p $\to$ n$\rho$}
\newcommand{\NstartoN}{\rm{N}^{\star} \rightarrow \rm{N}}
\newcommand{\rch}{$R_{\mathrm{C/H}}$}
\newcommand{\rchdeuxh}{R_{\mathrm{CH}_{2}/\mathrm{H}}}
\newcommand{\rcchdeux}{R_{\mathrm{C/CH}_{2}}}

\newcommand{\mevcc}{MeV/$c^{2}$}
\newcommand{\gevcc}{GeV/$c^{2}$}
\newcommand{\gevc}{GeV/$c$}
\newcommand{\mevc}{MeV/$c$}
\newcommand{\agev}{$A$\,GeV}
\newcommand{\Ekinmath}{E_{\mathrm{kin}}}

\begin{document}
\title{Inclusive \bf{ \epem } production in collisions of pions
with protons and nuclei in the second resonance region of baryons
}

\author{R.~Abou~Yassine$^{6,13}$, J.~Adamczewski-Musch$^{5}$, O.~Arnold$^{10,9}$, E.T.~Atomssa$^{13}$, 
M.~Becker$^{11}$, C.~Behnke$^{8}$, J.C.~Berger-Chen$^{10,9}$, A.~Blanco$^{1}$, C.~Blume$^{8}$, 
M.~B\"{o}hmer$^{10}$
% ...
}

\affiliation{
(HADES collaboration) \\\mbox{$^{1}$LIP-Laborat\'{o}rio de Instrumenta\c{c}\~{a}o e F\'{\i}sica Experimental de Part\'{\i}culas , 3004-516~Coimbra, Portugal}\\
\mbox{$^{2}$AGH University of Science and Technology, Faculty of Physics and Applied Computer Science, 30-059~Kraków, Poland}\\
\mbox{$^{3}$Institute of Nuclear Physics, Polish Academy of Sciences, 31342~Krak\'{o}w, Poland}\\
\mbox{$^{4}$Smoluchowski Institute of Physics, Jagiellonian University of Cracow, 30-059~Krak\'{o}w, Poland}\\
\mbox{$^{5}$GSI Helmholtzzentrum f\"{u}r Schwerionenforschung GmbH, 64291~Darmstadt, Germany}\\
\mbox{$^{6}$Technische Universit\"{a}t Darmstadt, 64289~Darmstadt, Germany}\\
\mbox{$^{7}$Institut f\"{u}r Strahlenphysik, Helmholtz-Zentrum Dresden-Rossendorf, 01314~Dresden, Germany}\\
\mbox{$^{8}$Institut f\"{u}r Kernphysik, Goethe-Universit\"{a}t, 60438 ~Frankfurt, Germany}\\
\mbox{$^{9}$Excellence Cluster 'Origin and Structure of the Universe' , 85748~Garching, Germany}\\
\mbox{$^{10}$Physik Department E62, Technische Universit\"{a}t M\"{u}nchen, 85748~Garching, Germany}\\
\mbox{$^{11}$II.Physikalisches Institut, Justus Liebig Universit\"{a}t Giessen, 35392~Giessen, Germany}\\
\mbox{$^{12}$Frederick University, 1036~Nicosia, Cyprus}\\
\mbox{$^{13}$Laboratoire de Physique des 2 infinis Irène Joliot-Curie, Université Paris-Saclay, CNRS-IN2P3. , F-91405~Orsay , France}\\
\mbox{$^{14}$Nuclear Physics Institute, The Czech Academy of Sciences, 25068~Rez, Czech Republic}\\
\mbox{$^{15}$LabCAF. F. F\'{\i}sica, Univ. de Santiago de Compostela, 15706~Santiago de Compostela, Spain}\\
\mbox{$^{16}$Uniwersytet Warszawski - Instytut Fizyki Do\'{s}wiadczalnej, 02-093~Warszawa, Poland}\\
\mbox{$^{17}$Warsaw University of Technology, 00-662~Warsaw, Poland}\\
\mbox{$^{18}$Bergische Universit\"{a}t Wuppertal, 42119~Wuppertal, Germany}\\ 
\\
\mbox{$^{a}$ also at Instituto Politécnico de Coimbra, Instituto Superior de Engenharia de Coimbra, 3030-199~Coimbra, Portugal}\\
\mbox{$^{b}$ also at Helmholtz Research Academy Hesse for FAIR (HFHF), Campus Darmstadt, 64390~Darmstadt, Germany}\\
\mbox{$^{c}$ also at Technische Universit\"{a}t Dresden, 01062~Dresden, Germany}\\
\mbox{$^{d}$ also at Helmholtz Research Academy Hesse for FAIR (HFHF), Campus Frankfurt , 60438~Frankfurt am Main, Germany}\\
\mbox{$^{e}$ also at Charles University, Faculty of Mathematics and Physics, 12116~Prague, Czech Republic}\\
\mbox{$^{f}$ also at Czech Technical University in Prague, 16000~Prague, Czech Republic}\\
\mbox{$^{g}$ also at Dipartimento di Fisica and INFN, Universit\`{a} di Torino, 10125~Torino, Italy}\\
\mbox{$^{h}$ also at University of Wroc{\l}aw, 50-204 ~Wroc{\l}aw, Poland}\\
}

\begin{abstract}
  Inclusive \epem\ production has been studied with HADES in \pimp, $\pi^-$ + C and $\pi^- + \mathrm{CH}_2$ reactions, using the GSI pion beam at $\sqrt{s_{\pi p}}$ = 1.49~GeV. Invariant mass and transverse momentum distributions have been measured and reveal contributions from Dalitz decays of $\pi^0$, $\eta$ mesons and baryon resonances. The transverse momentum distributions are very sensitive to the underlying kinematics of the various processes. The baryon contribution  exhibits a deviation up to a factor seven from the QED reference expected for the dielectron decay of a hypothetical point-like baryon with the production cross section constrained from the inverse $\gamma$ n$\rightarrow \pi^-$ p  reaction. The enhancement is attributed to a strong four-momentum squared dependence of the time-like electromagnetic transition form factors as suggested by Vector Meson Dominance (VMD). Two versions of the VMD, that differ in the photon-baryon coupling, have been applied in simulations and compared to data. VMD1 (or two-component  VMD) assumes a coupling via the $\rho$ meson and a direct coupling of the photon, while in VMD2 (or strict VMD) the coupling is only mediated via the $\rho$ meson. The VMD2 model, frequently used in transport calculations for dilepton decays, is found to overestimate the measured dielectron yields, while a good description of the data can be obtained with the VMD1 model assuming no phase difference between the two amplitudes. Similar descriptions have also been obtained using a time-like baryon transition form factor model where the pion cloud plays the major role.
\end{abstract}

\maketitle


\section{Introduction}
Radiative transitions of an excited baryon (R) to a nucleon (N) and a virtual massive photon ($\gamma^*$), converting to a dilepton pair (Dalitz decays), carry important information about the electromagnetic structure of the baryon in the time-like region. It is encoded in electromagnetic  Transition Form Factors (eTFF) that describe the baryon-photon interaction vertex. Magnetic, electric and Coulomb form factors are the most common representation of eTFFs used in the description of R $\to$ N$\gamma^*$ transitions.  Generally, form factors are analytic functions of the squared four momentum transfer \qdeux\ which, in the Dalitz decay process, is the dilepton mass $q^2=\mllmath ^2 $~\cite{Krivoruchenko:2002_Ann_Phys}. A strong enhancement of an eTFF as a function of increasing  $\mllmath $ is expected by the Vector Meson Dominance (VMD) as an effect of the intermediate light vector mesons $\rho/\omega/\phi$. Such an enhancement of the effective eTFF, defined as the sum of squares of the moduli of the form factors, has been observed by HADES for the first time for baryonic transitions in the second resonance region ($i.e.$ baryon masses around 1.5 \gevcc ) by means of pion-nucleon scattering  \cite{jointPRL}. There, the discussion of the results focused on the exclusive channel \pimptonee , the extraction of the effective eTFF and comparison of microscopic models for baryonic transitions to data.

In this paper, we extend the presentation of dielectron production to inclusive distributions obtained from $\pi^-+p$ and \pimC\ reactions. They provide an important reference for the understanding of dielectron production off nuclei and in heavy-ion collisions in the few-\agev\ energy region. Indeed, in such reactions decays of baryon resonances are of particular interest since they provide contributions from primary \NN\ and secondary \piN\ collisions. These contributions are dominant for invariant masses above the $\eta$ meson mass and need to be known before drawing conclusions on in-medium effects. For example, studies of dielectron production in   \pNb\ collisions at $\Ekinmath=3.5$~GeV revealed a significant cold matter effect, observed as an enhancement in the invariant mass distribution below the $\rho$ meson pole w.r.t. the \pp\ reference measured at the same energy~\cite{Agakishiev14_pp35_exclusive,HADES:2012sui}. One possible explanation of this effect was attributed to production and Dalitz decay of baryonic resonances. Such a mechanism was corroborated by microscopic transport calculations (GiBUU) assuming a two-step decay of baryonic resonances $R\rightarrow N\rho\rightarrow Ne^+e^-$~\cite{Weil12}. Also descriptions of dielectron production in heavy-ion collisions in other transport models include such processes~\cite{Bratkovskaya:2007jk, Weil12, Endres:2015fna}. 
The two-step decay scheme strongly relies on the applicability of VMD to baryon Dalitz decays. Moreover, it assumes  that the intermediate $\rho$ meson saturates the baryon-virtual photon transition amplitude (``strict VMD"). However, as pointed out by several authors \cite{Friman97,Krivoruchenko:2002_Ann_Phys,Zetenyi12},  such a strict VMD approach overestimates radiative decays of baryon resonances. Consequently, various extensions of VMD  including either excited vector mesons or a two-component photon coupling scheme, i.e.\ direct and via intermediate vector mesons, were suggested to cure this problem.  It is therefore important to test those approaches using an elementary process with sufficient sensitivity. 

In principle the VMD hypothesis can be directly tested for baryons by measuring Dalitz decays. 
They, however, are not easy to measure due to the low branching ratios ($BR \sim 10^{-5}$) and because of ambiguities in the assignment of the contributing resonances. 
The data obtained with the High Acceptance Di-Electron Spectrometer (HADES) in proton-proton reactions at $\Ekinmath =1.25$\,GeV, i.e.\ below the $\eta$ meson production threshold,  allowed to identify the $\Delta(1232)$ Dalitz decay~\cite{Hades17_DeltaDalitz}. 
The analysis indicates a mass dependence of the effective eTFF in line with VMD. 
Measured \epem\ invariant mass distributions show an enhancement of about a factor two at invariant masses $\meemath\approx 400$\,\mevcc\  w.r.t.\ predictions for point-like resonances, signaling an important contribution from off-shell $\rho$'s~\cite{Hades17_DeltaDalitz}. 
Measurements of \pp\ reactions at higher energy, $\Ekinmath=3.5$\,GeV, also indicated an enhancement at higher invariant masses, presumably due to contributions from higher mass resonances. 
However, determination of the individual resonance contributions depends both on the resonance production amplitudes and on the $\rho$ meson couplings to the various baryon transitions. 
Constraints on the resonant production amplitudes were obtained from one-pion production \cite{Agakishiev14_pp35_exclusive} but the isolation of individual transitions in \pp\ collisions was not possible because of the overlap of the various baryon resonances excited in the collision and since branching ratios to the $\rho$n channel are poorly known.

Pion-nucleon reactions offer a much better sensitivity. Baryon resonances with a  mass equal to $\sqrt{s_{\pi N}}$, can be excited in the s-channel and hence can be studied more directly. 
Indeed, recent HADES experiments with the GSI secondary pion beam allowed for a Partial Wave Analysis (PWA) with the Bonn-Gatchina framework (Bn-Ga) of the two-pion production in \pim p  in the second resonance region, around $\sqrt{s_{\pi p}}=1.49$~~GeV~\cite{HadesPionbeamTwopi}. 
This analysis allowed for the extraction of N(1440), N(1520) and N(1535) resonance contributions and the determination of the branching ratios for decays into various 2$\pi$ final states ($\Delta \pi, \sigma$ N, $\rho$N). 
The results for the $\rho$N decay were particularly important for calculating the dielectron invariant mass distributions and the extraction of effective eTFF of the resonances published in \cite{jointPRL}. 
As will be discussed below, this information is also crucial for the interpretation of the inclusive  and exclusive \epem\ production in context of the various VMD approaches mentioned above. We will investigate in particular two versions of VMD, 
the first one (VMD1 or two-component VMD) using a two-coupling ($\gamma$+$\rho$) scheme based on the $\gamma \rho$ Lagrangian of \cite{Kroll67} and the second  assuming baryon-virtual photon coupling only via $\rho$ based on the Lagrangian of \cite{Sakurai69} (VMD2 or "strict" VMD). These two versions  
were compared in theoretical studies of dielectron production in pion-induced reactions and brought about significantly different results~\cite{Effenberger:1999nn}. 
In a recent publication of the GiBUU group, calculations using VMD2 are presented and compared to the  data measured in the exclusive channel~\cite{Gallmeister:2022jco}.

To scrutinize these different approaches we present a comprehensive comparison of the two VMD versions to the inclusive and previously published exclusive data for the two collision systems \pimp\ and \pimC .  In addition to VMD models, we also use an  eTFF model which has been extended to the time-like region \cite{Ramalho17,Ramalho20}.
We also present details of the data-driven model calculations applied to the description of the exclusive \pimptonee\ channel~\cite{jointPRL}.  
As the data available for \epem\ emission in $\pi$N reactions is very scarce and limited to the region of the $\Delta(1232)$ resonance or below \cite{Berezhnev76_SovJourn24,Hoffman83,Jerusalimov18}, our data fill an important gap. 

The structure of our paper is the following:
Sec.~\ref{sec:set-up} and \ref{sec:DataAnalysis} describe the experimental set-up and data analysis procedure, Sec.~\ref{sec:Simulations} presents the ingredients of the simulations. 
Inclusive and exclusive dielectron distributions are compared to simulations in Sec.~\ref{sec:Results}, and conclusions are drawn in Sec.~\ref{sec:Conclusion}.
\section{Experimental set-up}
\label{sec:set-up}
A secondary pion beam with a central momentum of $p_{\mathrm{beam}}=0.685$\,\gevc\ was used.
On average $10^5$ $\pi^-$/s  with a momentum spread of $\pm1.7~\%$ were transported by the beam line up to the target~\cite{Hades17_pibeam}. 
The contamination, dominated by electrons from
external conversion, reaches  about $9.6\,\%$. 
A 4.6\,cm long polyethylene (\chdeux ) cylinder and a 3-fold segmented stack of carbon (C) foils, each with a diameter of $12$ mm and equally spaced over a distance of $4.6$ cm, were used in separate runs.
A diamond detector with an active area of  $14\times 14\,$mm$^{2}$, placed in front of the target, provided a start signal for time-of-flight measurements in HADES \cite{Agakichiev09_techn}.
A coincidence of the start signal with a multiplicty of hits ($\ge 2$) in HADES triggered the data acquisition. HADES  has excellent reconstruction capabilities  for  \epem\ pairs.
It consists of six identical sectors covering the full azimuthal and the polar angles from $18^{\circ}$ to $85^{\circ}$ with respect to the beam axis. 
Each sector contains a hadron-blind Ring Imaging CHerenkov (RICH) detector for electron identification, two sets of Mini-Drift Chambers (MDCs) and a time-of-flight wall (TOF) accompanied by a pre-shower detector at lower polar angles (18$^{\circ}$ to $45^{\circ}$). 
The time-of-flight (TOF) is measured by Resistive Plate Counters and scintillators covering polar angles smaller and larger than $45^{\circ}$, respectively.
Momentum vectors of produced particles are calculated from the deflection of tracks measured in the MDCs, with one set in front of and the other one behind a magnetic-field region provided by the six coils of the super-conducting toroid.  
The four-momentum vectors of the positron (e$^+$) and electron (e$^-$) are reconstructed with a precision of $1-2~\%$.
Electron identification is provided by combining the signals measured by the RICH, the tracking and TOF systems. 
More details on particle identification (PID) and tracking in HADES can be found in~\cite{Agakichiev09_techn, Sellheim_PhD}.
\section{Data analysis}
\label{sec:DataAnalysis}
Electron pair candidates (like-sign and opposite-sign) were selected by an algorithm which requested at least two fully reconstructed tracks in the MDCs, assigned velocities measured in TOF of $v/c > 0.9$ and momenta larger than $100$\,\mevc . 
Those tracks were extrapolated back to determine the intersection points with the spherical mirror of the RICH~\cite{Agakishiev12_pp22}.
The ring width ($\sigma$) was estimated and calculated as a function of the primary vertex position, the polar angle and
the azimuthal angle. 
Finally, a region of interest (ROI) was defined to search for fired pads on the RICH pad plane by smearing the ring radius as $r = r_0 \pm 3 \sigma$ around the position derived from the intersection points. 
At least three pads inside ROI were requested to qualify the track as electron candidate. 
Furthermore, it was demanded that the ``point'' of closest approach of the electron track to the beam axis was located inside the target and only one hit is found in the start detector.
More details on the e$^{+}$ and e$^{-}$ reconstruction in this experiment can be found in~\cite{Scozzi18,Rodriguez-Ramos2021}. 
 
The combinatorial background (CB) is estimated bin-by-bin using the geometrical average of the numbers of like-sign pairs reconstructed in the same event: $N_{\mathrm{CB}}=2 \sqrt{N_{\rm{e}^+\rm{e}^+}N_{\rm{e}^-\rm{e}^-}}$. 
If for a given bin (of the invariant mass or transverse momentum) either $N_{\rm{e}^+\rm{e}^+}$ or $N_{\rm{e}^-\rm{e}^-}$ is zero, the arithmetic sum is used:
$N_{\mathrm{CB}} = N_{\rm{e}^+\rm{e}^+}+N_{\rm{e}^-\rm{e}^-}$.

\begin{figure}[h]
\includegraphics[width=0.48\textwidth]{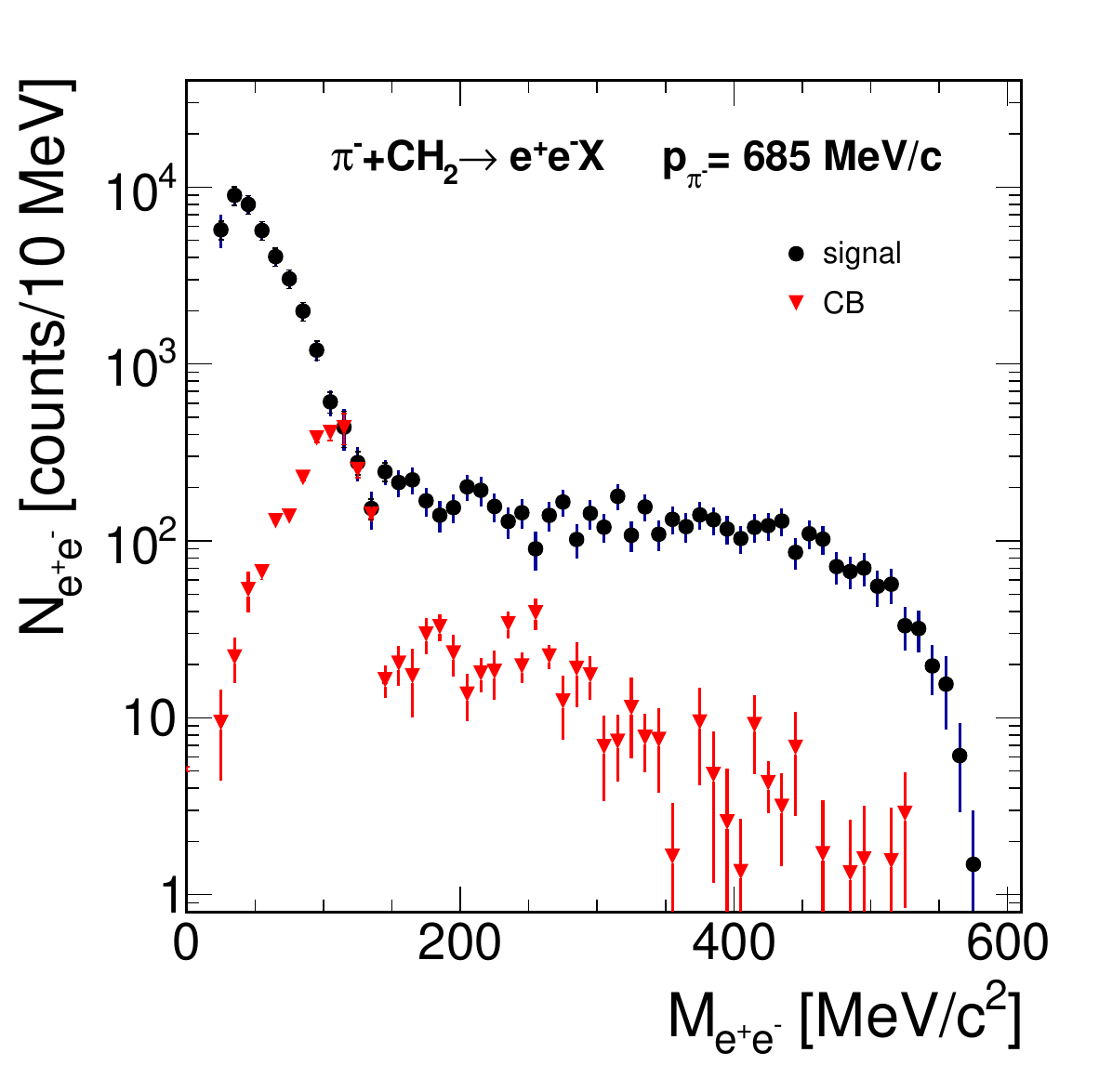} 
\caption{Invariant mass distribution of signal pairs (black discs) and combinatorial background (CB, red triangles)  measured in the HADES acceptance and corrected for detection efficiency. Statistical errors are shown as vertical bars. 
}
\label{fig:CB}
\end{figure}
 In Fig.~\ref{fig:CB}, the CB distribution as a function of the invariant mass is compared to the correlated signal obtained after CB subtraction. It can be clearly observed that the contribution of the CB to the total \epem\ yield is significant only around 120\,\mevcc , where a strong contribution of correlated background arises from double-conversion of both photons from the \piz\ decay.   
In total, 16 340 signal \epem\ pairs have been reconstructed in reactions on the \chdeux\ target. 
To derive distributions of dielectrons from \pimp\ reactions in the polyethylene target, the contribution from pion interactions with carbon needs to be subtracted. 
This was achieved by separately measuring the interactions of pions with the carbon target. 
The respective normalization factor for carbon data was estimated using high statistics events with pions from measurements on the carbon and polyethylene targets. 
The dedicated subtraction procedure is described in detail in \cite{HadesPionbeamTwopi}.  
\begin{figure}[h]
\includegraphics[width=0.48\textwidth]{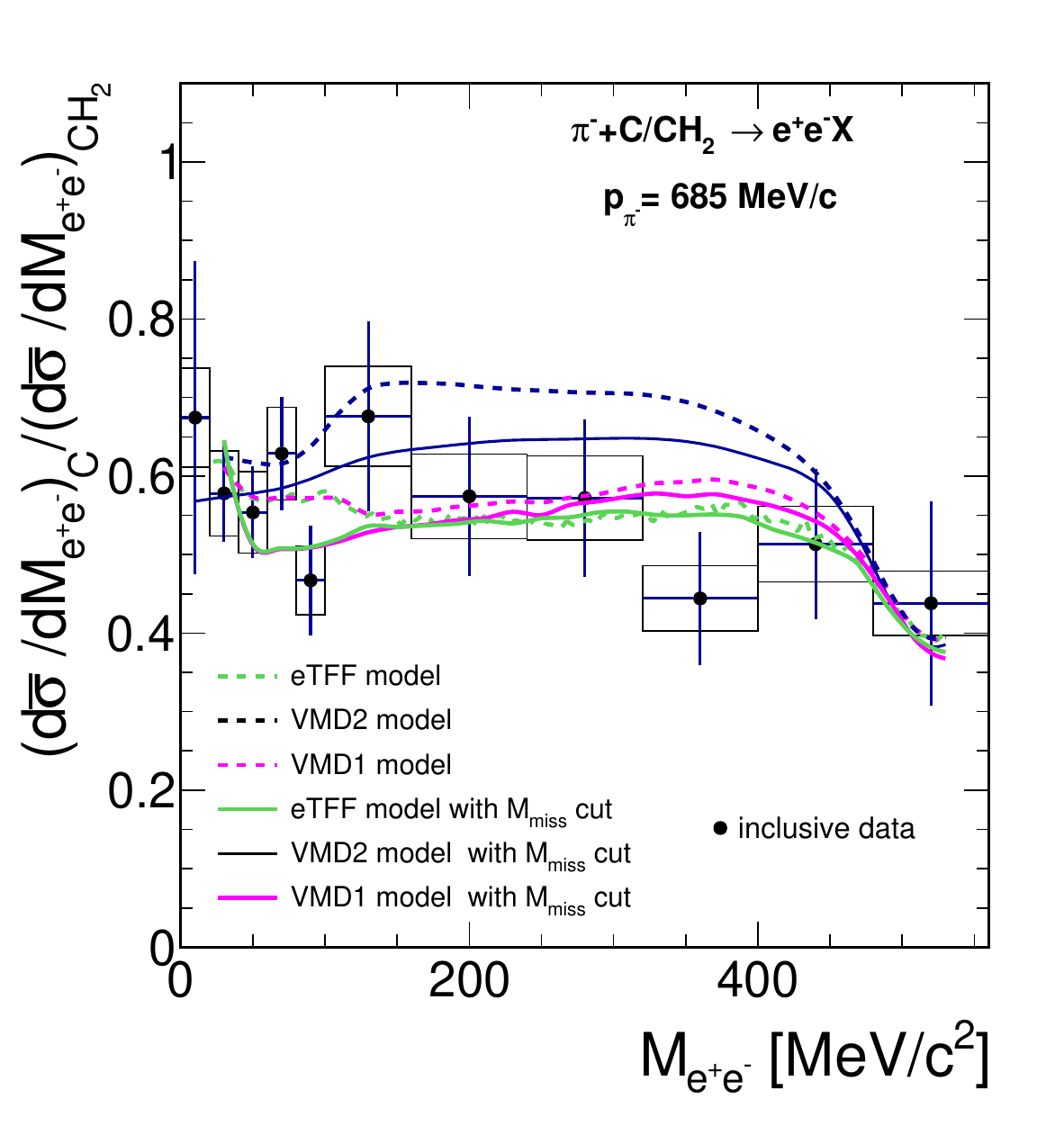}
\caption{Ratio of \epem\ production cross sections  measured on the carbon to polyethylene targets (black discs) as a function of \mee .
The data is for the inclusive analysis and show statistical (vertical lines) and systematic uncertainties (boxes). The horizontal width reflects the bin width. 
The various curves display the results of \epem\  simulations for the  eTFF (green), VMD2  (black) and VMD1 (magenta) models without (dotted lines) and with the missing mass cut (full lines).
For details of the simulations see Sec.~\ref{sec:Simulations}.}
\label{fig:CPEratio}
\end{figure}

The ratio $\left(\mathrm{d}\bar{\sigma}/\mathrm{d}\meemath\right)_{\rm{C}}/\left(\mathrm{d}\bar{\sigma}/\mathrm{d}\meemath\right)_{\rm{CH_2}}$ between the inclusive dielectron production cross sections for reactions on CH$_2$ and on C is displayed in Fig.~\ref{fig:CPEratio} as a function of the invariant mass. Here and throughout the text, $\bar{\sigma}$ denotes cross sections integrated over the HADES acceptance.
The ratio of inclusive yields each integrated over the whole invariant mass range is 
$\rcchdeux = 0.59  \pm 0.02 $ (stat) $\pm$ 0.01 (syst), where the dominant systematic error results from the relative normalization of the carbon and polyethylene measurements. 
This value corresponds to a ratio of \pimC\ to \pimp\ yields \rch =   $2 R_{\mathrm{C/CH2}}/(1- R_{\mathrm{C/CH2}}) = 2.88 \pm 0.18 $ (stat) $\pm 0.12$ (syst). 
It should be stressed that the main contribution to this averaged ratio comes from the region below \mee = 140~\mevcc, which corresponds to the \piz\ Dalitz decay. 
However, no significant dependence of the ratio is observed as a function of invariant mass (see Fig.~\ref{fig:CPEratio}). 
The ratios based on  the simulations are discussed in Sec.~\ref{sec:Simulations}.

The reconstructed \pimp\ elastic scattering yields $N_\text{el}$ of known cross sections $\sigma_\text{el}$ (from SAID) were used to calculate the integrated accepted luminosity (see for details \cite{HadesPionbeamTwopi}). 
The cross sections for the dielectron production on CH$_2$, proton and $\mathrm{C}$ (obtained from CH$_2$ after subtraction of the contribution from proton interactions), are given inside the HADES acceptance after correction for the efficiency accounting for losses in electron-positron detection and reconstruction. 
The efficiency correction factors $\epsilon (p,\theta,\phi$) were calculated as 3-dimensional functions of the electron momentum, the polar and the azimuthal angles.
These efficiencies are understood as detection efficiency, as they include corrections for tracks which were in the acceptance, defined by $acc(p,\theta,\phi$) (see below), but did not traverse the active area of all tracking and PID detectors used for reconstruction.
Electron tracks were processed with GEANT simulations of the HADES detection system and reconstructed using the same algorithm as employed for the experimental data analysis. 
These simulations are also used to calculate resolution effects on the reconstructed electron momenta. 
They are particularly important for modeling the \epem\ missing mass distribution, presented in~\cite{jointPRL}, used for the exclusive channel separation.
The electron momentum resolution is parameterized as a function of the true  momentum (GEANT input) and the emission angles. 
These momentum smearing functions together with the acceptance matrices $acc(p,\theta,\phi$), calculated separately for positrons and electrons, constitute the HADES filter. 
The filter, available on request, allows to compare any model to the data measured in the HADES acceptance. 
\section{Simulation inputs}
\label{sec:Simulations}
 In the collision-energy range investigated with our experiment, several  sources of \epem\ pairs are expected to contribute and were accordingly implemented in simulations using the PLUTO event generator \cite{Froehlich07}. 
 The most important are (i) Dalitz decays of neutral pions (\piz $\to \gamma$ \epem\  with $BR=1.2~\%$), originating from single or double production (ii) $\eta$ Dalitz decays ($\eta \to \gamma $\epem ) with $BR=6.9\times 10^{-3}$ and (iii) exclusive \pimptonee\ reactions with contributions from the  R$\rightarrow$ \epem n (Dalitz decay of point-like baryon) and $\rho \, (\rightarrow$ \epem ) n  final states. These two contributions are described in sub-sections A and B, respectively.  

For the pion production cross sections we use data from earlier measurements and the SAID data base (\cite{SAIDdatabase,Bulos64,Starostin05,Prakhov04}). 
The excitation function of $\eta$ close to the production threshold was measured in~\cite{Prakhov05} and included in our simulations.
Less important are channels with $\Delta$ Dalitz decays in $\Delta^{+,0}\pi^{-,0}$ final states (with $BR=4.2\cdot 10^{-5}$). The respective production cross section were derived from the PWA analysis of the two-pion production \cite{HadesPionbeamTwopi}. 
The individual contributions of these channels is about two orders of magnitude smaller than the $\eta$ channel and was skipped in simulations.
Table~I summarizes the production cross sections of various final states applied in the simulations.   
\begin{table}[ht]
    \centering
  \caption{List of  cross sections for various  final states in $\pi^-$p collisions.}
  \label{tab:channels}
  \begin{tabular}{|c|c|}
    \hline
     Final state & $\sigma$ [$ mb$] \\
    \hline
    \hline
    \piz n & $9.0\pm 0.8$ \\
    \piz \piz n & $1.9\pm 0.1$ \\
    \pim \piz p & $4.0\pm 0.5$ \\
     $\eta$ n & $0.63\pm 0.2$ \\
    $\Delta^+ \pi^-$ & $0.78\pm 0.08$ \\
    $\Delta^0 \pi^0$ & $1.0\pm 0.1$ \\
    $\rho$ n & $1.35\pm 0.20$ \\
    n$\gamma $ & $0.22$ \\
    \hline
  \end{tabular}
\end{table}

To estimate dielectron production in \pimCHdeux\ collisions, the reaction channels listed above must also be calculated for \pimC\ collisions. 
In our simulations they were modeled using a spectator model described in Sec.~\ref{sec:sim-C}.  
\subsection{{\boldmath $\pi^- $}  p {\boldmath$\rightarrow$ } R {\boldmath $\rightarrow$ n e$^+$e$^-$}: QED reference}
The baryon Dalitz decays were modeled using results of the PWA of Bonn-Gatchina (Bn-Ga) group on $\gamma$\,n $\to \pi^-$\,p and $\pi^-\,$p $\rightarrow \pi\pi\,$N reactions. 
As the PWA of both $\gamma$ n and $\pi\pi$ final states in the \pimp\ reaction  indicate that the dominant resonant contributions are s-channel N(1520) ($J^P$= 3/2$^-$) and N(1535) ($J^P$= 1/2$^-$) \cite{HadesPionbeamTwopi} formation, our calculation of dielectron production was conducted for Dalitz decays of these states
We used detailed balance to calculate the cross section of the inverse $\pi^{-}\mathrm{p} \rightarrow \gamma$ n  reaction ($\sigma({\pi^- \mathrm{p} \rightarrow \mathrm{n} \gamma}) = 220$~$\mu$b). 
 Assuming the formation of a baryon resonance (N$^*$) with a  given spin and parity in the \pimp\ entrance channel, the differential cross section for its Dalitz decay ($\mathrm{N}^* \rightarrow $ n\epem ) was calculated as a function of the invariant mass using equations derived in \cite{Krivoruchenko:2002_Ann_Phys}: 
\begin{eqnarray}
\frac{\mathrm{d}\Gamma^{\NstartoN \epemmath}}{\mathrm{d}\meemath } 
= \frac{2\alpha}{3\pi \meemath } \Gamma^{\NstartoN \gamma^*}(\meemath ).
\label{eq|dalitz3}
\end{eqnarray}
$\Gamma^{\NstartoN  \gamma^*}(\meemath)$ is related to the radiative decay width  $\Gamma^{\NstartoN \gamma}$ as follows:
\begin{eqnarray}
\Gamma^{\NstartoN \gamma^*}(q^2)=\Gamma^{\NstartoN \gamma} \frac{\sigma_{+}^{3}\sigma_{-}}{m_{+}^{3}m_{-}} \frac{\left|G_T(q^2)\right|^2}{\left|G_T(0)^2\right|^2},
\label{eq:RadDal}	
\end{eqnarray}
where $m_{\pm} = M_\mathrm{R} \pm M_\mathrm{N}$ and $\sigma_{\pm}^2  =  m_{\pm}^2-q^2$. 
$M_{\mathrm{R}}$ is the resonance mass, $M_{\mathrm{N}}$ the nucleon mass and $\alpha$ the fine-structure constant. 
The form factor  $\left|G_T\right|$ is a combination of the squared moduli of the electric ($G_{E}$), magnetic ($G_M$) and Coulomb ($G_C$) form factors and defined by
 
\begin{align}\label{eq:effectiveonehalf}
 \left|G_T (q^2) \right|^2 &= 2\left|G_E (q^2)\right|^2  +\frac{q^2}{M_{\rm{R}}^2}\left| G_C(q^2)\right|^2
 \end{align}
for $J^P=1/2^-$, and
\begin{align}
\begin{split}
\left|G_T(q^2)\right|^2 =& \left|G_{E} (q^2)\right|^2+3\left|G_{M}(q^2)\right|^2  \\
&+\frac{q^2}{2M_{\rm{R}}^2}\left|G_C (q^2) \right|^2
\label{eq:effectivethreehalf}
\end{split}
\end{align}
for $J^P = 3/2^-$.
We note that Eq.~(2) is strictly valid  for both $J^P=1/2^-$ and 3/2$^-$.  Other $J^P$ configurations would lead to steeper mass distributions (for details, see \cite{Krivoruchenko:2002_Ann_Phys}).  

The electric, magnetic and Coulomb form factors are commonly used by models of eTFFs which provide their mass dependence. 
In case of simulations based on the eTFF model of \cite{Ramalho16,Ramalho17,Ramalho20} we have used the respective form factors taken from calculations for the N(1520) and N(1535) transitions, which have a very similar invariant mass dependence. 
However, in the limit of point-like baryons, representing in our simulations the QED reference, we have used covariant form factors $g_1,~g_2$ and $~g_3$. 
These form factors are directly connected to the baryon-photon interaction vertex and can be assumed constant while $G_E,~G_M$ and  $G_C$ are related by transformations which depend on the dielectron invariant mass. 
The respective relations are given  by equations  (III.18/19) and (III.24/25) in \cite{Krivoruchenko:2002_Ann_Phys}. 
A careful analysis of the dependence of the Dalitz decay width on the covariant form factors was presented in \cite{Zetenyi03_formula} (see Fig.~1 there). 
It shows that for $J^P=1/2^-$ and $J^P=3/2^-$ transitions the dominant contributions originate from terms related to $g_1$ and $g_2$, respectively. 
Moreover, in the latter case these two contributions are of very similar shape. 
Interference effects between the contributions of various isospins might also affect the invariant mass distribution of pairs but are negligible in our case.
Therefore, in calculations we have used just one form factor. 
 
The integrated cross section for the \pimptonee\ reaction is sensitive to the differential cross section at small \mee , which does not depend strongly on the $J^P$ configurations. 
Hence, the Dalitz decay width at small \mee\ can be approximated as
\begin{align}\label{eq:Dalitzbranching}
\Gamma_{\mathrm{QED}}^{\NstartoN \epemmath}\approx\frac{2\alpha }{3\pi} \Gamma^{\NstartoN \gamma}  \log \bigg( \frac{m_{-}}{2m_{\rm{e}}} \bigg) ,\\
\intertext{where  $m_{\rm{e}}$ is the electron mass. For\  $M_{\rm{R}}\approx$~1.49  \gevcc }
 \Gamma_{\mathrm{QED}}^{\NstartoN \epemmath }\sim 1.34\ \alpha \ \Gamma^{\NstartoN \gamma }. 
\end{align}
A similar relation can be derived for the cross sections
\begin{eqnarray}
\sigma({\pi^- \rm{p} \rightarrow \rm{N} \epemmath })\approx 1.34\ \alpha \ \sigma({\pi^- \rm{p} \rightarrow \rm{n} \gamma}).
\label{eq:Dalitzsigma}
\end{eqnarray}

Finally, with $\sigma({\pi^- \mathrm{p} \to \mathrm{n} \gamma }) = 220~\mu$b, we obtain $\sigma({\pi^- \mathrm{p} \to \mathrm{n} \epemmath }) = 2.1\,\mu$b as normalization for the Dalitz decays.
Taking into account this cross section value and Eq.~(\ref{eq|dalitz3}) for the differential distribution as a function of \mee , a well defined reference (``QED") is obtained, which is in agreement with measurements of the $\pi^- \mathrm{p} \, \to$ \, n$\gamma$ cross section.
In the case of the eTFF model, the simulations were normalized,   at small invariant masses, where the effect of mass dependence of the form factors is negligible, to the same values as the QED simulations, i.e.\ according to Eq.~(\ref{eq:Dalitzbranching}). 
\subsection{\boldmath$\pi^- \rm{p}\to R \to n \rho (\to \epemmath $): VMD models}
\label{sec:sim-B}
In this section we discuss the two-step decay process using the two VMD versions introduced above. 
Both approaches take into account the $\rho$ decay but with different expressions for the differential decay widths, as described below. 
The mass distribution of the produced $\rho$ is constrained by the results of the PWA of two-pion production  \cite{HadesPionbeamTwopi}. 
\subsubsection{$\rho$ spectral function and partial decay widths}
The $\rho$ meson spectral function has the form
\begin{eqnarray}
 \mathcal{A}(M) \sim \frac{M^2\Gamma_\text{tot}(M)}{(\mrhomath ^2 - M^2)^2 + M^2\Gamma^2_\text{tot}(M)},
\label{eq:BW2}
\end{eqnarray}
where $M$ is the $\rho$ mass and \mrho\ is its pole value. 
The total width $\Gamma_\text{tot}(M)$ is the sum of the  $\rho \to $ \pippim\ and $\rho \to $ \epem\ partial decay widths:
 \begin{eqnarray}
\Gamma_\text{tot}(M) = \Gamma_{\pippimmath}(M) + \Gamma_{\epemmath}(M).
\label{eq:gammatot}
\end{eqnarray}
Here, it is assumed that Eq.~(\ref{eq:BW2}) describes the $\rho$-meson spectral function also below the two-pion threshold. 
The spectral function, as defined in Eq.~(\ref{eq:BW2}), derives from the relation between the total width and the imaginary part of the self-energy $\Sigma_{\rho}(M)$:   \begin{eqnarray}
\operatorname{Im}(\Sigma_{\rho}(M))= M \Gamma_\text{tot}(M).  
\label{eq:SE}
\end{eqnarray}
To be consistent with the Bn-Ga PWA approach, we used the following expressions of the decay widths :
\begin{eqnarray}
\Gamma_{\pippimmath}(M)= \Gamma_{0} \left(\frac{\mrhomath}{M}\right)^2 \left(\frac{q_{\pi\pi}}{q_{0,\pi\pi}}\right)^3\frac{1+R^2q_{0,\pi\pi}^2}{1+R^2q_{\pi\pi}^2}, \label{eq:BW1pipi}
\label{eq:pipiwidth}
\end{eqnarray}
where $q_{\pi\pi}=\frac{1}{2}(M^2-4m_{\pi}^2)^{1/2}$ is the center-of-mass momentum for the two-pion decay of the $\rho$ meson with the running mass $M$  and $q_{0,\pi\pi}$ is the value for $M=\mrhomath$. 
The cut-off parameter is $R=0.8\,$fm and the width at the pole of the resonance is $\Gamma_{0}$. 
Equation~(\ref{eq:pipiwidth}) is similar to the Manley parameterization, commonly used in transport approaches, which, however, uses a softer cut-off value of $R=1\,$fm \cite{Gallmeister:2022jco}. 

For the $\rho\to$ \epem\ decay, we use the 
VMD1 (two-coupling scheme) and VMD2 (strict) models, as defined in \cite{Oconnell95}, which are  based on two different Lagrangians for the $\gamma$-hadron coupling. 
In VMD1 it is described by two terms representing a direct $\gamma$ and $\gamma-\rho$ transition with independent coupling constants. 
For baryon resonances the coupling constants can be fixed by fitting the radiative and the N$\rho$ decay widths separately. 
In the case of VMD2  the transition is only mediated via $\rho$ mesons. 
The parts of the two Lagrangians describing the $\rho-\gamma$ coupling are given by  
\begin{eqnarray}
\mathcal{L}_{\rho \gamma}^\text{VMD1}=\frac{-e}{2g_{\rho}}F^{\mu\nu}\rho^0_{\mu\nu},\
\mathcal{L}_{\rho \gamma}^\text{VMD2}=\frac{-eM_{\rho}^2}{g_{\rho}}\rho^0_{\mu}A^\mu ,
\label{eq:VDM2}
\label{eq:VDM1}
\end{eqnarray}
 where F$^{\mu\nu}$ and $\rho^0_{\mu\nu}$ are the electromagnetic and $\rho$ meson field tensors, $A^\mu$ is the photon field, $g_{\rho}$ the dimensionless $\rho - \gamma$ coupling constant  ($g_\rho$=5.96) and $e$ the electron charge.
The respective partial decay widths  are given by
\begin{eqnarray}
 \mathrm{VMD1: \ }  &\Gamma(\meemath)_\text{VMD1} = &\Gamma_{0} \ \meemath / M_{\rho}  \label{eq:VMD1} \\ 
\mathrm{VMD2: \ }  & \Gamma(\meemath)_\text{VMD2} = &\Gamma_{0} \left( M_{\rho} / \meemath \right)^3. 
\label{eq:VMD2}
\end{eqnarray}
We wish to stress that these expressions for the partial widths derive from the choice of total width. 
If the relation $\operatorname{Im}(\Sigma_{\rho})= \mrhomath \Gamma_\text{tot}(M)$ is used instead of Eq.(~\ref{eq:SE}), then the spectral function reads:
\begin{eqnarray}
 \mathcal{A}(M) \sim  \frac{\mrhomath^2\Gamma_\text{tot}(M)}{(\mrhomath ^2 - M^2)^2 + \mrhomath^2\Gamma^2_{tot}(M)}
\label{eq:BW1}
\end{eqnarray}
and the expressions of the partial decay widths differ by a factor $M$/\mrho \ w.r.t.\ Eqs.~(\ref{eq:pipiwidth}),  (\ref{eq:VMD1}) and (\ref{eq:VMD2}).
Simulations of the \pimptonee\ and \pimptonpippim\ were performed using  a modified version of the Pluto event generator \cite{Froehlich07} with expressions for the decay widths given in Eqs.~(\ref{eq:pipiwidth}),  (\ref{eq:VMD1}) and (\ref{eq:VMD2}). 
\subsubsection{$\mathbf{\rho}$ production in $\pi^-p$ collisions}
\begin{figure}[hbt] 
\includegraphics[width=0.48\textwidth]{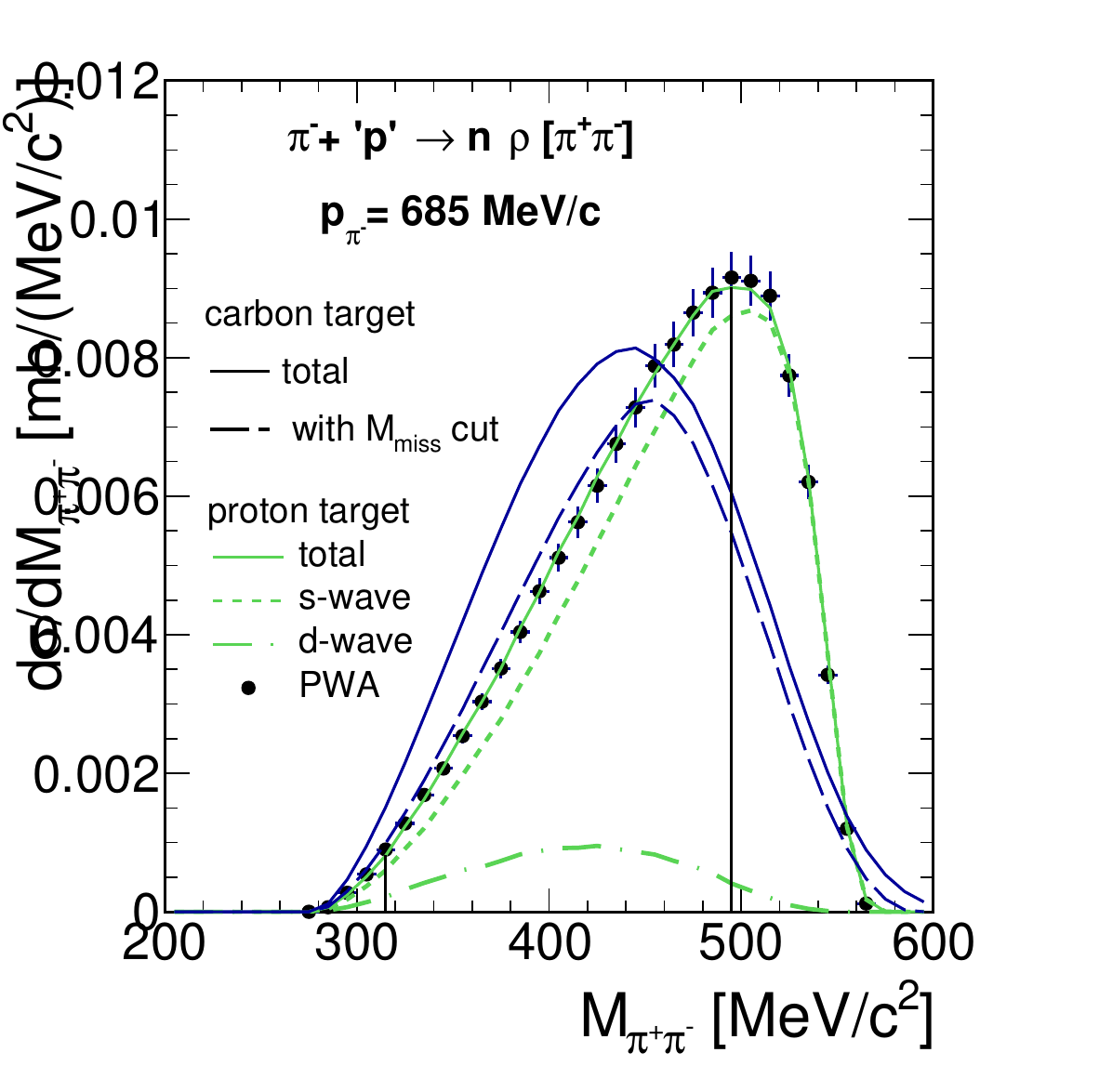}
		\caption{Invariant mass distribution of pion pairs from $\rho$ meson decay in \pimptonpippim\ reaction obtained from the PWA (black circles)\cite{HadesPionbeamTwopi}. Vertical bars account for the systematic uncertainties of the analysis (statistical errors are negligible). The curves display the result of a fit using simulated distributions for s- (dashed green) and d-wave (dash-dotted green) $\rho$-n production. The sum  is shown as solid green curve. The vertical lines show the range in invariant mass used for the fit. The solid black curve displays the total $\rho$ mass distribution for the \pimC\ case normalized to the same $\rho$ production cross section as for the \pimp\  reaction. For the dashed black curve, only $\rho$ mesons selected by the \epem\ analysis, with the missing mass cut 900~$<$ \mmiss  (\mevcc) $<$ 1030 , are considered.} 		
		\label{fig:fitpipi_rho}
		\end{figure}
  
\begin{figure}[hbt]
	  \includegraphics[width=0.48\textwidth]{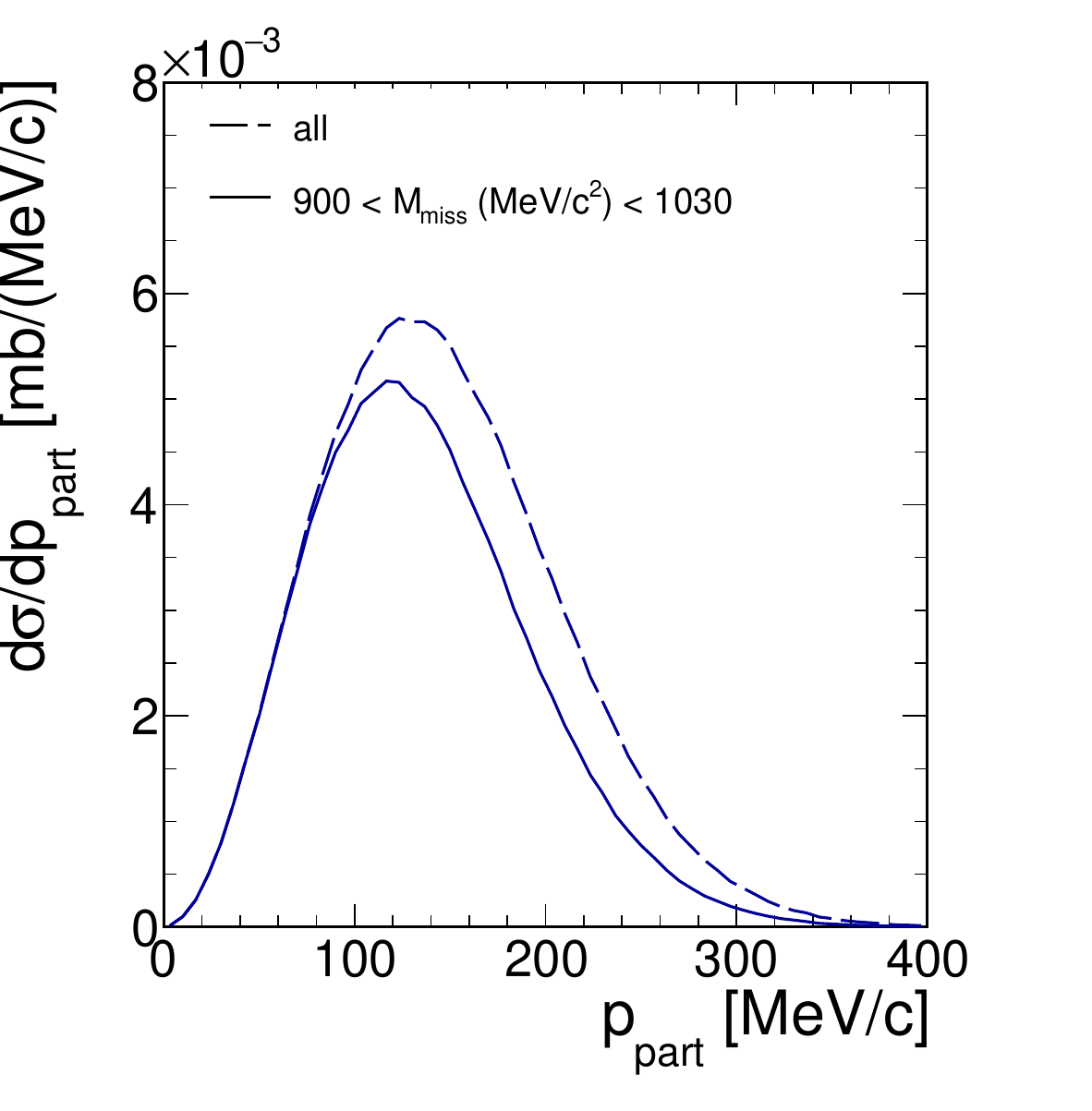}	
	 \caption{The distribution of the participant proton momentum 
   for the quasi-free \pimp $\to$ n$\rho$ reaction in carbon is shown for all events (dashed curve) and after the mising mass selection 900~$<$ \mmiss  (\mevcc) $<$ 1030 (solid curve). The spectrum is normalized to the free \pimpnrho\ cross section.}		
		\label{fig:fitpipi_Fermi}
		\end{figure}
A realistic description of the $\rho$ meson mass distribution in the \pimp\ reaction is essential for a consistent prediction of the \pimptonee\ differential cross sections. 
In the simulations, the mass distribution of produced $\rho$ mesons was fixed to the one obtained in the PWA for the \pimptonpippim\ reaction at the same energy \cite{HadesPionbeamTwopi}. 
According to the PWA, the total cross section for this channel amounts to $\sigma_{\rho}=1.35 \pm 0.12$\,mb and is dominated by s-channels with about $50~\%$ contribution from N(1520). 
The corresponding differential cross section is displayed as black circles in Fig.~\ref{fig:fitpipi_rho}, where the errors represent the systematic uncertainties of the analysis. 
We stress that this $\rho$ meson mass distribution comprises only the low mass tail of the spectral function given by Eq.~(\ref{eq:BW2}) as a consequence of the restricted phase space for $\rho$ production in \pimpnrho\ at $\sqrt{s_{\pi p}}=1.49$\,GeV. 
For the calculation of the total width, the  $\rho\,\to\,$~\epem\ partial decay width was chosen following the VMD2 model (Eq.~(\ref{eq:VMD2})). However, the actual choice of the distribution function affects the \pippim\ production yield only very close to the threshold.
Furthermore, the $\rho$ meson mass distribution is affected by the available phase space in the reaction and depends on the orbital momentum in the $\rho-$N system: for s-wave  production no dependence on the $\rho$ momentum in the center-of-mass of the \pimpnrho\ reaction ($p_\text{CM}$) is expected while, for d-wave production, events were further weighted by a factor $p_\text{CM}^2$. 
The invariant mass distributions from PWA were used to determine the relative contribution of d-wave (dashed dotted green curve) to s-wave (dashed green curve) by means of a fit, as shown in Fig.~\ref{fig:fitpipi_rho}. 
The fit  was performed in the range of invariant masses [320, 490]~\mevcc  in order to limit the sensitivity to resolution effects or to the incident pion beam momentum distribution. 
The best agreement with the PWA solution is obtained  with a 10$\pm$ 2~\% d-wave contribution.   
\subsubsection{$e^+e^-$ distributions}
The $\rho$  production model described above was used to generate  \epem\ events  from the $\rho$ decay with two options for the dielectron partial decay widths: VMD1 or VMD2 (given in Eqs.~(\ref{eq:VMD1}) and (\ref{eq:VMD2}), respectively). 
The VMD2 model provides by itself a description of the full \pimptonee\ process. 
In the case of VMD1, the vanishing $\rho\to~$\epem\ partial decay width for \mee $\rightarrow 0$ is compensated by a contribution corresponding to the coupling of the photon to point-like baryons. 
The latter is given by the QED reference introduced above.  
As the phase between the $\rho$ and $\gamma$ amplitudes is unknown, we fixed it to have the maximum yield, to better describe the data. 
The yields  generated with the VMD1 $\rho$ decay model and with the QED reference models, respectively, were therefore simply added. 
\subsection{Modelling of quasi-free \boldmath$\pi^-+p$ reactions in  the carbon nucleus}
\label{sec:sim-C}
\begin{figure}[hbt]  	
    \includegraphics[width=0.45\textwidth]{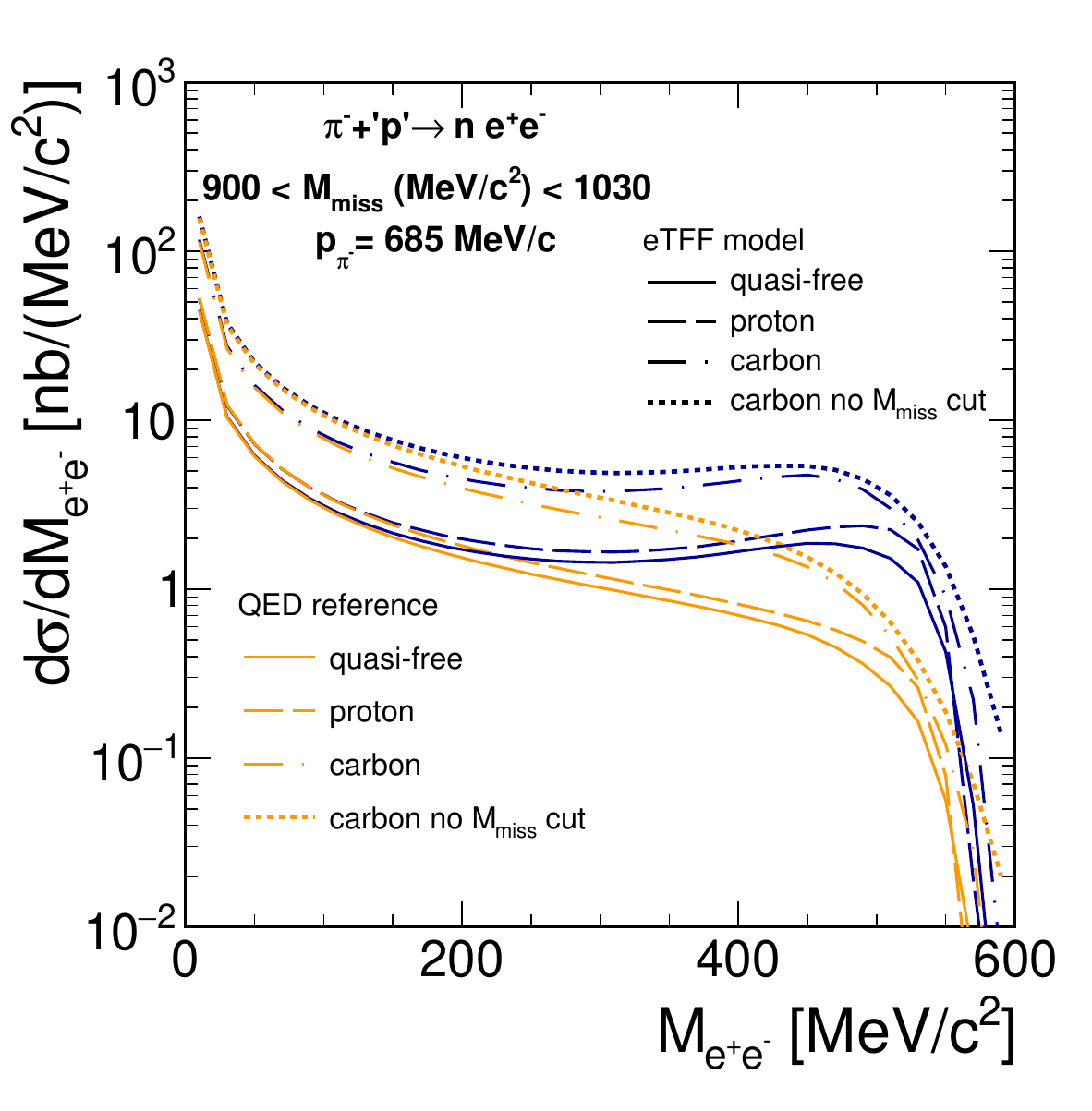} 
		\caption{Dielectron differential production cross sections as a function of the \epem\ invariant mass obtained from simulations of baryon Dalitz decay are shown in the full solid angle for the \pimp\ reaction (long dashed line), for the \pimC\ reaction with  900~$<$ \mmiss  (\mevcc) $<$ 1030 (dash-dotted line) and for the quasi-free reaction (normalized by $Z_\text{eff}$, solid line). The results obtained for the \pimC\ reaction without the missing mass cut are shown for comparison (dotted lines). The orange curves correspond to the eTFF model and the black curves display the QED reference.}
		\label{fig:Minv_proton_carbon_etff}
		\end{figure}

\begin{figure}[h]
	\centering
\includegraphics[width=0.45\textwidth]{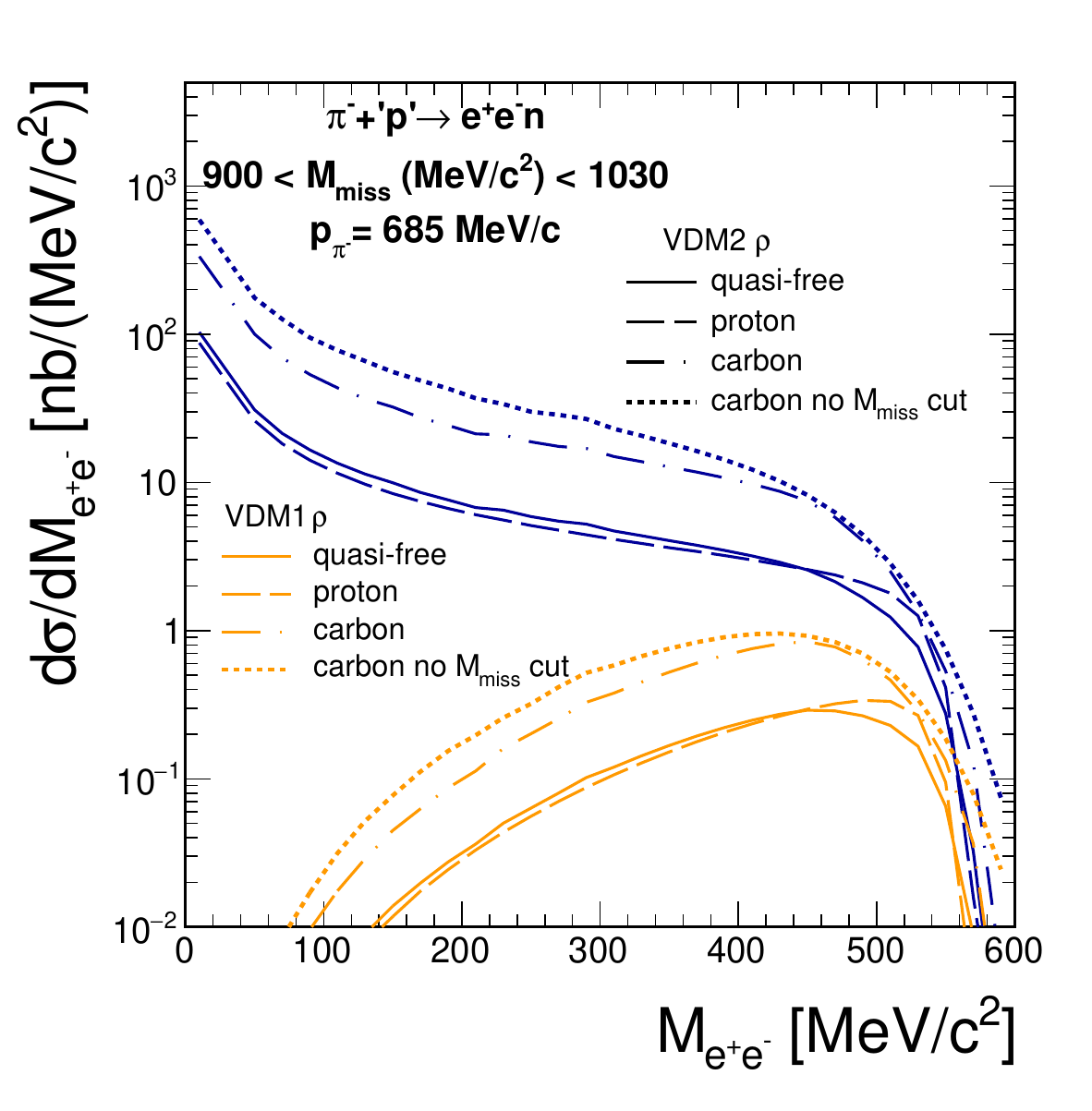}\\
		\caption{ Same as Fig.~\ref{fig:Minv_proton_carbon_etff} but for the $\rho$ contribution calculated with VMD2 (black curves) and VMD1 (orange curves) models.}
		\label{fig:Minv_proton_carbon_VDM}
\end{figure}

To describe the \epem\ production off the carbon and the polyethylene targets, interactions with carbon nuclei were modeled using a quasi-free participant spectator model, $i.e$ 
 the $^{12}\mathrm{C}$ nucleus is represented as a spectator on-shell $^{11}\mathrm{B}$ and a participant off-shell proton. 
The proton momentum distribution is taken into account in PLUTO as obtained from quasi-free $^{12}\mathrm{C}(e,e'p)$ measurements~\cite{Nakamura76} and is displayed in Fig.~\ref{fig:fitpipi_Fermi}.  
A condition on the  \epem\ missing mass 900~$<$ \mmiss  (\mevcc) $<$ 1030 w.r.t.\ pion-nucleon system was applied in the analysis to select the quasi-free \pimptonee\ exclusive production channel (see Fig.~1 of~\cite{jointPRL}). 
This cut reduces the contribution of large proton momenta in the carbon nucleus. 
In addition, it suppresses the contribution of mechanisms beyond the quasi-free approach (inelastic processes, rescatterings, short range correlations, ...).
        
For the $\eta$ production off $^{12}\mathrm{C}$, the steep increase of the cross section as a function of the energy above the production threshold has been taken into account within the participant-spectator model. 
It results in a similar average cross section for \pimp\ interactions in $^{12}\mathrm{C}$ as for the p target (see \cite{Ramstein:2020jkr} for details).
In addition to the scaled yields, the \epem\ distributions obtained for $^{12}\mathrm{C}$ are affected by the difference in phase space, due to the $p$ momentum in $^{12}\mathrm{C}$. 
For the $\eta$ and \piz\ Dalitz decays and for the QED reference and eTFF model, this effect just distorts the high invariant mass region, as shown for the two latter models in Fig.~\ref{fig:Minv_proton_carbon_etff}. 

The case of the VMD models is different. 
Indeed, the $\rho$ mass distribution is significantly affected by the difference in  available energy in the \pimp\ center-of-mass for the carbon target, which results from the participant spectator model and favors $\rho$ mesons with lower masses, as shown in Fig.~\ref{fig:fitpipi_rho}. 
Due to this effect, the \epem\ yields in the region below  400\,\mevcc\ are increased by a factor of 6.1 (instead of 2.9) - see Fig.~\ref{fig:Minv_proton_carbon_VDM} (dotted lines). 
The effect of the missing mass cut, which is used for the analysis to select the exclusive quasi free \pimptonee\ channel, is also illustrated in Fig.~\ref{fig:Minv_proton_carbon_VDM} (dash-dotted lines). 
It can be observed that the cut reduces the difference between the shapes of the distributions in proton and carbon.  
This is due to the fact that the distortion of the invariant mass distribution is mostly due to the large proton momenta, which are suppressed by the cut, as shown in Fig.~\ref{fig:fitpipi_Fermi}.

Differential  \epem\ cross sections on the polyethylene targets are deduced from the corresponding ones on p and C:
\begin{eqnarray}
    \left( \frac{d\sigma}{d\meemath} \right)_{\mathrm{CH}_2} = \left( \frac{d\sigma}{d\meemath}\right)_{\mathrm{C}} + 2 \left(\frac{d\sigma}{d\meemath}\right)_{\mathrm{H}} .       
        \label{eq:CH2}
        \end{eqnarray}
         To deduce from the \chdeux\ measurements  a cross section comparable to a free cross section, we define a "quasi-free" cross section as:
        
    \begin{align}
    \left(\frac{\mathrm{d}\sigma}{\mathrm{d}\meemath}\right)_{qf}   =  \left(\frac{\mathrm{d}\sigma}{\mathrm{d}\meemath}\right)_{\mathrm{CH}_2} / \rchdeuxh   \quad,
        \end{align}
        with
         $\rchdeuxh = 2 + R_{\mathrm{C/H}}$, following from Eq.(\ref{eq:CH2}).
       \noindent
Taking $\rcchdeux$ = 0.59, as found experimentally, one obtains \rch = 2.9 and  $\rchdeuxh$ = 4.9.
\begin{figure}[h!bt]
\includegraphics[width=0.41\textwidth,left]{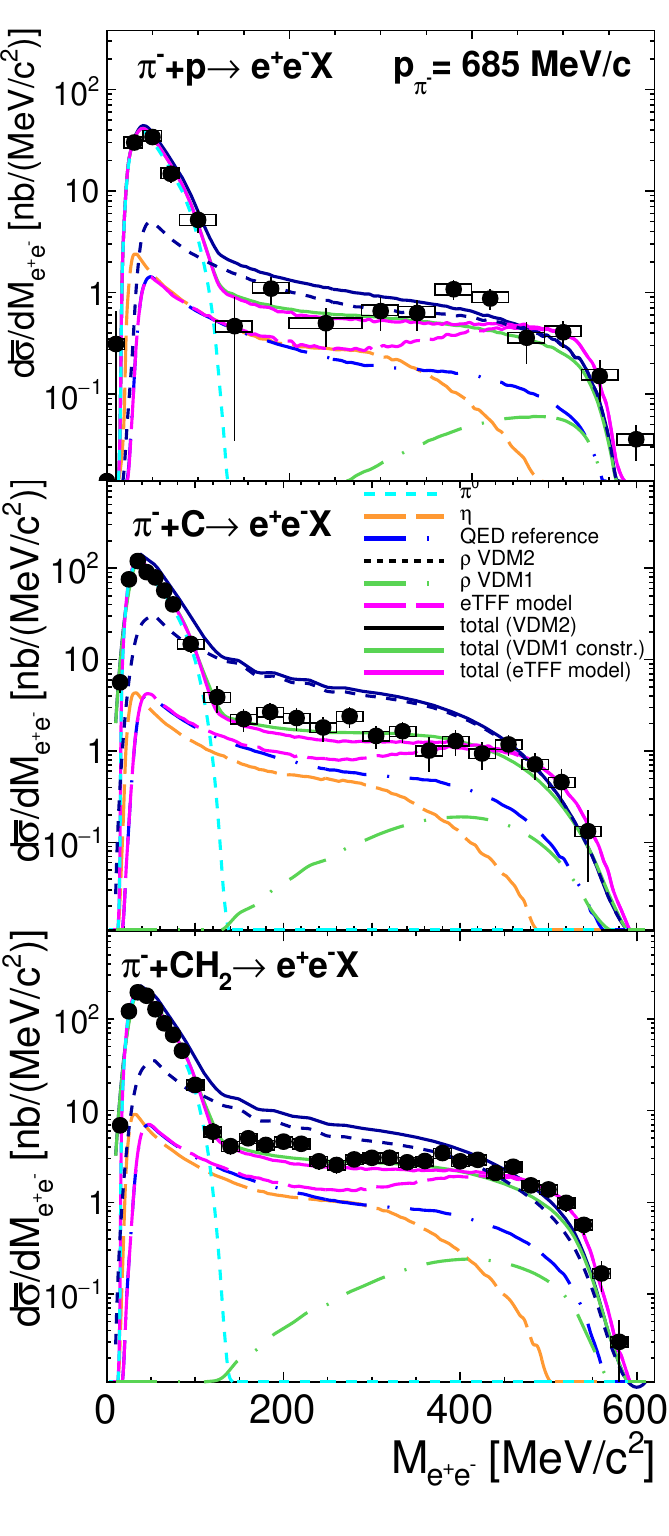} 
 
  \caption{Inclusive differential cross sections, in the HADES acceptance and corrected for efficiency, as a function of the \epem\ invariant mass for \pimp\ , \pimC\  and \pimCHdeux\ reactions are shown by symbols. Total and systematic uncertainties are shown by bars and boxes, respectively. The black, green and magenta solid curves display the results of the simulations for the "VMD2", "VMD1 constr." and "eTFF model", which use the same \piz\ and $\eta$ contributions (dashed cyan and dashed orange curves, respectively). Contributions from $\rho$ decay are shown separately for  VMD2 (black dashed) and VMD1 (green dot-dashed).  VMD1 constr. is a coherent sum of the QED reference (dot-dashed blue) and $\rho$ decay. For the eTFF model, the contribution of the baryon Dalitz decay with eTFF (dashed magenta) is also displayed.}
  \label{fig:InvMass_3targets}
\end{figure}
\begin{figure*}[hbt]
\includegraphics[width=0.9\textwidth]{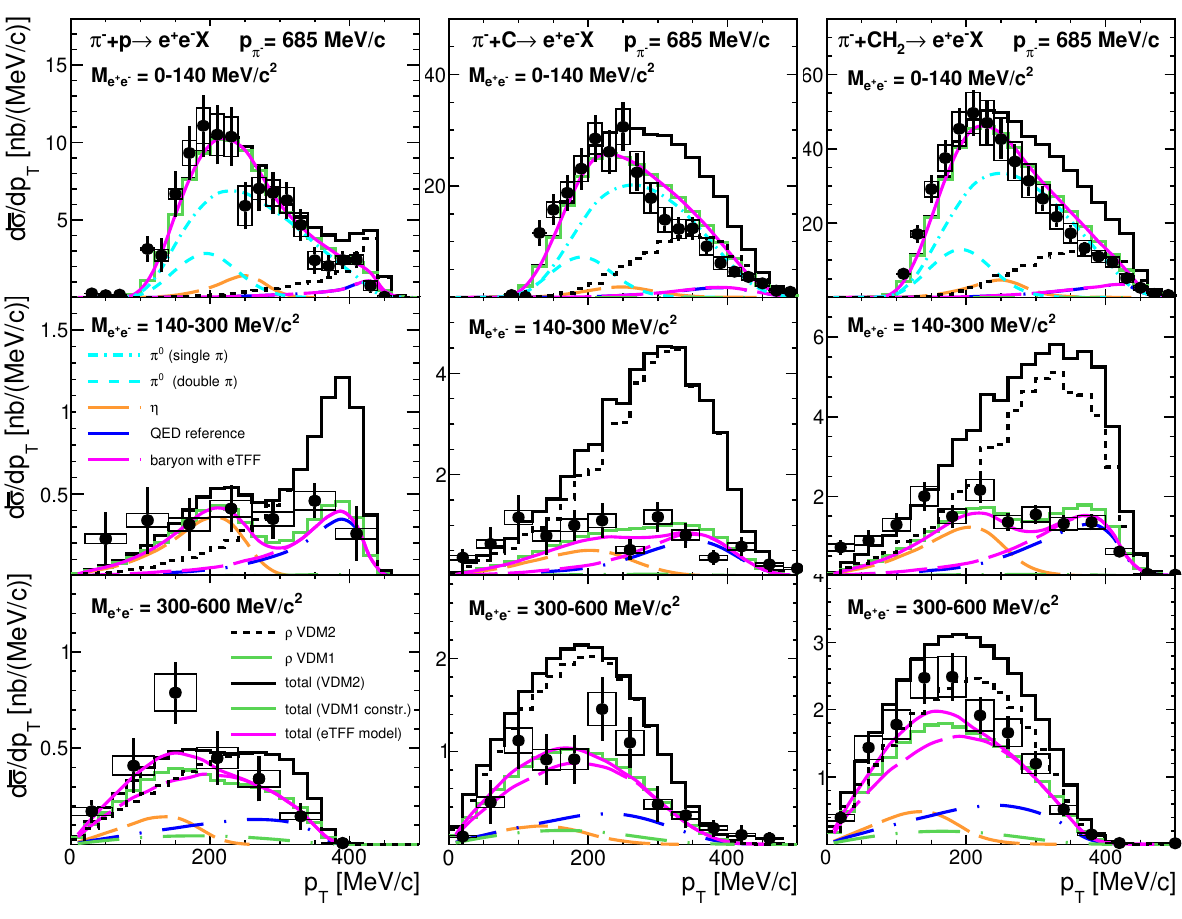} 
\caption{
Inclusive differential cross sections as a function of the transverse momentum $p_{\rm{T}}$ of the \epem\ pair for \pimp\  (first column), \pimC\ (second column) and \pimCHdeux\ (third column) reactions are shown by symbols in the HADES acceptance after efficiency corrections. Total and systematic errors are shown by bars and boxes, respectively. Pairs with \mee\ $< $140~\mevcc, 140 $<$ \mee\ (\mevcc)$<$  300, and  300 $<$ \mee\ (\mevcc) $<$  600 are selected.  The color code for the simulation curves is as in Fig.~\ref{fig:InvMass_3targets}, except for the \piz\ contribution, which is shown separately for the one and two-pion processes as cyan dash-dotted and dashed lines, respectively.}
\label{fig:Pt_pi0}
\end{figure*}
The distributions for the quasi-free process were calculated for the \chdeux\ target within the missing mass cut (900~$<$ \mmiss  (\mevcc) $<$ 1030 ) and normalized by $Z_{\rm{eff}}+2=4.9$. As one can see in Figs.~\ref{fig:Minv_proton_carbon_etff} and \ref{fig:Minv_proton_carbon_VDM}, the quasi-free distributions (solid lines) are very similar to the ones for the \pimp\ reaction (long dashed), except for invariant masses above 400~\mevcc\ where the aforementioned effects of high momentum tails in the proton momentum distribution in carbon is still visible. Below this value, the shapes of the distribution are very similar and the difference is about 20~\%.
The ratios of the simulated dielectron differential yields from C and \chdeux\ targets are shown in Fig.~\ref{fig:CPEratio} for the eTFF, VMD2 and VMD1 models in comparison to the data.  Solid  lines  represent the ratio for the inclusive distributions, while the dotted lines show the ratio after the missing mass cut. One can note that the missing mass cut reduces the ratio for the VMD2. The higher value of the ratio for the VMD2 model is due to the phase space effect, as discussed above. 
\section{Results}
\label{sec:Results}
\subsection{Inclusive distributions}
Dielectron differential cross sections in the HADES acceptance as a function of the \epem\ invariant mass for \pimp\ (top panel), \pimC\ (middle panel) and \pimCHdeux\ (bottom panel) reactions are shown by symbols in Fig.~\ref{fig:InvMass_3targets}. 
Total and systematic errors are shown by vertical bars and boxes, respectively.  
For the $\rho$ decay, the two versions of VMD were considered in our simulations and are compared separately to the data: VMD1 (dot-dashed green) and VMD2 (dashed black). 
In case of VMD1, the dielectron mass distribution is vanishing at the two-pion mass threshold while in the other one  it is strongly increasing with the decreasing invariant mass according to the $1/M^3$ dependence of the partial decay width.
The data are compared to simulations described above and to the eTFF model of \cite{Ramalho16,Ramalho17} (dashed magenta) with added contributions from the meson Dalitz decays (solid magenta).
The QED reference representing the contribution expected for point-like baryons is shown by a dot-dashed blue curve. 
As expected, it is smoothly converging to the eTFF model at small masses. 
The $\pi^0$ mass region is well described in all cases, mainly by contributions originating from single and double pion production (cyan dashed), except for the calculation with VMD2, where some overshoot is visible. 
This overshoot is clearly apparent in the mass region above the pion mass.  
In the case of VMD1 (solid green), the coherent sum  of the QED reference (point-like baryon, dot-dashed blue)  and VMD1 $\rho$ (dot-dashed green) contributions with constructive interference describes the data very well.

Differential cross sections as a function of the transverse momentum of the \epem\ pair are shown in Figs.~\ref{fig:Pt_pi0}, for \pim + p (first column), \pim + C (second column) and \pimCHdeux\ (third column) reactions for three bins of the invariant  mass: \mee $<$ 140~\mevcc, $140<$\mee (\mevcc ) $<300$ and   300 $<$ \mee (~\mevcc ) $<  600$,  respectively. 
Below the pion mass, the distributions are dominated by the \piz\ Dalitz decay contribution, which is well described in our simulations. 
However, it is interesting to note that the data show already in this region a  good sensitivity to the exclusive process \pimptonee\ which contributes at the highest \pt. 
A clear overshoot  is observed  for the VMD2 model, while  the data are fairly well reproduced by the VMD1 (coherent sum) and eTFF models. 

In the intermediate mass region, the distributions are broader. The simulations clearly depict two structures, due to the $\eta$ Dalitz decay (orange dashed line)  and the exclusive \pimptonee\ process at the lowest and highest \pt, respectively. 
The $\eta$ contribution, which is particularly sensitive to the pion beam momentum spread in the case of the \pimp\ interactions and to the proton momentum distribution in the case of \pimC\ interactions, is satisfactorily described in the simulation for the three reactions. 
For the highest \pt, where the exclusive process dominates, the difference between the models is striking. 
The VMD2 model overshoots the yield in the high \pt\ region by about a factor 3, while the other models provide a much better description of the data. 
The calculation for the point-like baryon Dalitz decay (blue dot-dashed line) is also shown in the picture as a reference. 
It clearly shows that, for 140 $<$ \mee\ (\mevcc)$<$  300 and high \pt , the most important ingredient of these models is not the  eTFF, which has a small effect,  but the constraints provided by the \pim p $\to$ n$\gamma$ cross sections.  
One should also note some excess of the yield  above models at low \pt\ which might be explained by the \pimptonee\ bremsstrahlung, not included in the simulations.

In the highest invariant mass bin, which is mostly due to the exclusive \pimptonee process, the three models are much closer. 
The overshoot of the data by the VMD2 model is reduced, while the VMD1 and eTFF models are a bit below the data, so that the quality of the description of the data is similar in all models for this invariant mass region. 
In this region, which is closest to the $\rho$ meson pole, albeit still very much off-shell, the data are strongly in excess with respect to the point-like calculation. 
This points to the strong effect of the eTFF and VM contribution which is described in a realistic way in the three models, as discussed in more details below.  
\subsection{Exclusive \boldmath$\pi^-p\rightarrow e^+e^- n $ distributions.}
\begin{figure*}[hbt]
	\centering
\includegraphics[width=0.9\textwidth]{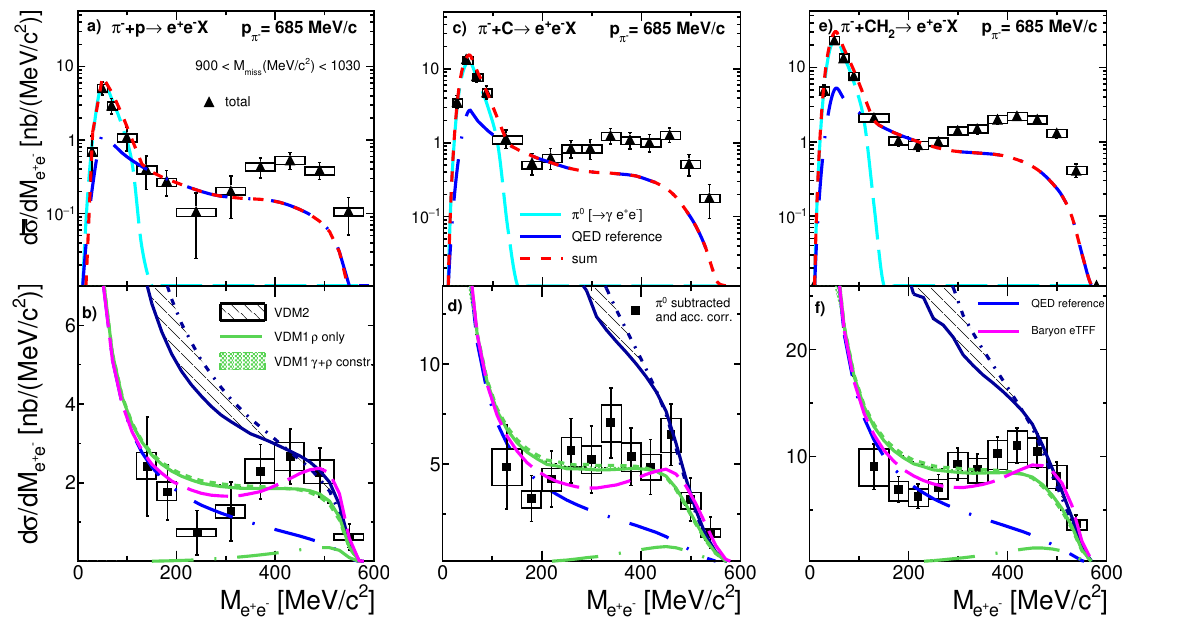}
    \caption{ 
      Dielectron invariant mass distributions for the \pimptonee X reaction integrated over the pair missing mass range [900-1030]\,\mevcc\ for the free (first column) and quasi-free reactions on C (second column) and \chdeux\ (third column).  First row: ( a), c) and e)):  Full triangles show experimental yields (d$\bar{\sigma}/$d$\meemath$)   integrated in the HADES acceptance with total and systematic errors displayed as bars and boxes.  The curves show the simulations for point-like baryon Dalitz decay (``QED reference", blue dash-dotted curve),   \piz\ Dalitz decay (cyan dashed curve) and the sum (red dashed curve). Second row : (b), d) and f)):  Data (full squares) are shown after subtraction of the \piz\ Dalitz decay contribution and acceptance corrections. Blue dot-dashed curve: QED reference, black dashed area: VMD2 model with d-wave contribution varied from 0 (full curve) to 10~\% (dashed curve), green full curve:  same for VMD1 models with constructive  sum of $\rho$ and $\gamma$ contributions, dot-dashed green curve: $\rho$ contribution from the VMD1 model. The effect of varying the d-wave contribution between 0 and 10\% is negligible for VMD1. Calculations using the  time-like eTFF model (magenta dashed curve)  are also shown. 
		}
	\label{fig:Minv}
\end{figure*}
The results on the exclusive dielectron production channel $n e^+e^-$ have been reported in \cite{jointPRL} for the \chdeux\ case. 
Here we recall the main findings and present comparisons to the two VMD models and the eTFF model for the proton and carbon case separately. 
The selection of the exclusive channel is achieved by the cut on the dielectron missing mass 900~$<$ \mmiss\  (\mevcc) $<$ 1030. 
For invariant masses \mee\ $>\,M_{\pi^0}$, a clear peak centered around the neutron mass is visible in the reactions on the proton and carbon (see Fig.~2 in \cite{jointPRL}). 
The mass distributions are very well described by simulations with the eTFF model and the two-component model. 
In particular, the description of the neutron peak on carbon provides evidence for the quasi-free production.  
However, at smaller invariant masses, the \piz\ Dalitz decay contribution is not fully suppressed by the missing mass cut. 
This is very clearly seen for the three reactions in the first row of Fig.~\ref{fig:Minv}. 
This contribution, being well described by the simulation, can be subtracted to further select the free or the quasi-free \pimptonee reactions. 
The corresponding data points are displayed for the three reactions \pimp , \pimC\  and \pimCHdeux\  in Fig.~\ref{fig:Minv} (lower row) and compared to the predictions of the two VMD versions.
The VMD2 version is found to give a reasonable description of yields measured for the highest invariant masses, but strongly overestimates the yields below 350~\mevcc. 
VMD2 overshoots the data in the low mass region and also does not provide the right slope. 
On the other hand, the two-component model with a constructive interference between amplitudes with $\rho$ and direct $\gamma$ couplings (full green curve)  provides a very good description.  
The hatched areas  visualises the effect of the $\rho$-n d-wave contribution, which is mostly significant for the VMD2 model. 
 
Conclusions from the exclusive channel are consistent with the one obtained from the analysis of the inclusive channel. 
It can indeed be observed that the application of the VMD2 approach for the dielectron production at low invariant masses strongly overestimates the production yields. 
This observation clearly points to the limitation of the "strict VMD" model in the baryon sector, previously suggested by some theoretical considerations, which can here be confirmed experimentally, thanks to the sensitivity offered by pion induced reactions. 
Recent calculations based on VMD2 \cite{Gallmeister:2022jco} show a similar slope as a function of the invariant mass, but a lower yield, due to a lower cross section for the $\rho$ production.   

\section{Conclusions}
\label{sec:Conclusion}
We presented results on  inclusive \epem\ production in \pimp , \pimC\ and \pimCHdeux\ reactions in the second resonance region which complement the study of the exclusive \epem\ production discussed in \cite{jointPRL} for the \chdeux\ target. 
The differential distributions as a function of the invariant mass and transverse momentum measured for the inclusive reaction within the HADES acceptance clearly show the contributions from Dalitz decays of neutral meson ($\pi^0$, $\eta$) and baryonic resonances. 
The former can be well described by using the known elementary cross sections, Transition Form-Factors for the meson decays and assuming a quasi-free process for the production off carbon nuclei, with a number of participant protons close to 3. 
In particular, the transverse momentum distributions are very sensitive to the different kinematics of the various processes. 
The invariant mass distributions indicate that the contribution from baryon resonances strongly deviates from a point-like behavior and highlights the contribution from off-shell $\rho$ mesons. 
This effect was taken into account in the simulations  within two implementations of the Vector Meson Dominance model: VMD1, which assumes a twofold photon-hadron coupling scheme, both direct and via $\rho$, and VMD2 which uses only the $\rho$ meson coupling. 
It was shown by detailed simulation studies that these two scenarios lead to very different predictions for the dielectron production. In the calculations, 
the PWA of $\rho$ meson production in \pimptonpippim\  provided a very important constraint. The comparison of  the simulations to the data reveals, for the first time, the importance of VMD in the description of Dalitz decays in the baryonic sector. 
The calculations assuming only direct photon coupling to point-like baryons, representing the QED reference, strongly underestimate the dielectron production in the mass region above the pion mass. 
However, a coherent sum of the QED reference and the $\rho$ meson contribution calculated within the VMD1 approach describes the invariant mass and transverse momentum distributions very well. 
Similarly do the  calculations based on microscopic models of baryon electromagnetic Transition Form Factors (etFF) \cite{Ramalho17,Ramalho20} with a photon coupling to a quark core and a pion cloud, where the latter plays a dominant role. 
On the contrary, the VMD2 model, assuming the saturation of the transition via intermediate $\rho$ mesons strongly overestimates the measured dielectron yields.  

The present study fully confirms  the  conclusions of  \cite{jointPRL} on the eTFFs derived from the studies of quasi-free \pimp\  reactions on \chdeux\ target  and extends them to the \pimp\ and \pimC\  systems. 
As discussed in \cite{jointPRL},  a missing mass selection allows to select the exclusive \epem\ production on a quasi free proton, which provides a better sensitivity to study eTFFs and VDM models. 
We have presented acceptance corrected exclusive \epem\ invariant mass distributions  and compared to the two VMD models. The data are well  described by the VMD1 approach, in contrast to VMD2 which strongly overshoots the measured dielectron yields. 
These results call for microscopic approaches taking into account constraints from both measurements of $\rho$ production in pion-nucleon reactions and  pion photoproduction, as discussed in \cite{jointPRL}.

These results constitute a unique basis of information for \epem\ production processes in the second resonance region. They provide important constraints for the model calculations of dielectron production in elementary and heavy-ion collisions and shed light on the electromagnetic structure of baryon transitions in the time-like region.

%
\begin{acknowledgments}
We appreciated the theoretical tools provided by A. Sarantsev, G. Ramalho and T. Pe\~na. 
 We are also very grateful for the interesting
discussions on theoretical aspects with K. Gallmeister and U. Mosel.
We acknowledge support from SIP JUC Cracow, Cracow (Poland), National Science Center, 2016/23/P/ST2/040 POLONEZ, 2017/25/N/ST2/00580, 2017/26/M/ST2/00600; TU Darmstadt, Darmstadt (Germany), DFG GRK 2128, DFG CRC-TR 211, BMBF:05P18RDFC1, HFHF, ELEMENTS:500/10.006, VH-NG-823,GSI F\&E, ExtreMe Matter Institute EMMI at GSI Darmstadt; Goethe-University, Frankfurt (Germany), BMBF:05P12RFGHJ, GSIF\&E, HIC for FAIR (LOEWE), ExtreMe Matter Institute EMMI at GSI Darmstadt; TU München, Garching (Germany), MLL München, DFG EClust 153, GSI TMLRG1316F, BmBF 05P15WOFCA, SFB 1258, DFG FAB898/2-2; JLU Giessen, Giessen (Germany), BMBF:05P12RGGHM; IJCLab Orsay, Orsay (France), CNRS/IN2P3, P2IO Labex, France; NPI CAS, Rez, Rez (Czech Republic), MSMT LM2018112, LTT17003, MSMT OP VVV CZ.02.1.01/0.0/0.0/18 046/0016066.
\\
The following colleagues from Russian institutes did contribute to the results presented in this publication but are not listed as authors following the decision of the HADES Collaboration Board on March 23, 2022: A. Belyaev, O. Fateev, M. Golubeva, F. Guber, A. Ierusalimov, A. Ivashkin, A. Kurepin, A. Kurilkin, P. Kurilkin, V. Ladygin, A. Lebedev, M. Mamaev, S. Morozov, O. Petukhov, A. Reshetin, A. Sadovsky, A. Shabanov and A.~Taranenko.
\end{acknowledgments}
%

\bibliography{dilepton_pionBeam}
\end{document}